\documentclass[reprint,twocolumn,eqsecnum,amsmath,amssymb,aps,nofootinbib,superscriptaddress,prx]{revtex4-1}
\usepackage{graphicx}
\usepackage{bm}
\usepackage{epsfig}
\usepackage{amssymb}
\usepackage{amsfonts}
\usepackage{amsmath}
\usepackage{color}
\usepackage{xcolor}
\usepackage{epstopdf}
\usepackage{hyperref}
\usepackage{float}
\usepackage{appendix}
\usepackage{wasysym}
\restylefloat{table}
\usepackage{bibentry}
\usepackage{multirow}
\usepackage[caption=false]{subfig}
\usepackage{braket}
\usepackage{bbm}
\newcommand{\ba}{\begin{eqnarray}}
\usepackage{manfnt}
\newcommand{\ea}{\end{eqnarray}}
\newcommand{\bd}{\begin{displaymath}}

\newcommand{\nn}{\nonumber \\}
\newcommand{\md}{~{\rm mod}~}

\DeclareGraphicsExtensions{.pdf,.png,.jpg}

\begin{document}
\title{Effective Field Theory of Dipolar Braiding Statistics in Two Dimensions}
\author{Yun-Tak Oh}
\affiliation{Department of Physics, Korea Advanced Institute of Science and Technology, Daejeon 34141, Republic of Korea}
\affiliation{Division of Display and Semiconductor Physics, Korea University, Sejong 30019, Korea}
\author{Jintae Kim}
\affiliation{Department of Physics, Sungkyunkwan University, Suwon 16419, Korea}
\author{Jung Hoon Han}
\email[Electronic address:$~~$]{hanjemme@gmail.com}
\affiliation{Department of Physics, Sungkyunkwan University, Suwon 16419, Korea}
\begin{abstract}
A rank-2 toric code (R2TC) Hamiltonian in two dimensions can be constructed as a Higgsed descendant of rank-2 U(1) lattice gauge theory. As noted by the authors recently, [Y.-T. Oh, J. Kim, E.-G. Moon, and J. H. Han, Phys. Rev. B {\bf 105}, 045128] the quasiparticles in that model shows unusual braiding statistics that depends on the initial locations of the particles which participate in the braiding. We show that this new kind of statistical phase captures the total dipole moment of quasiparticles encompassed in the braiding, in contrast to the conventional anyonic braiding seeing the total charge. An Aharonov-Bohm interpretation of such {\it dipolar braiding statistics} is made in terms of emergent, rank-1 vector potentials that are built out of the underlying rank-2 gauge fields. Pertinent field theories of the quasiparticle dynamics in the R2TC are developed, and the accompanying conservation laws derived. A {\it dipolar BF theory} of the rank-2 gauge fields is constructed and shown to correctly capture the dipolar braiding statistics, in contrast to the conventional BF theory capturing the monopolar braiding statistics of anyons in the rank-1 toric code. 
\end{abstract}
\date{\today}

\maketitle
\section{Introduction}
The Chern-Simons (CS) field theory~\cite{deser82} made its entry into condensed matter physics as an effective field theory of the fractional quantum Hall system~\cite{ZHK89,zhang92} and encodes the mutual statistics of anyonic quasiparticles~\cite{wilczek82a,wilczek82b}. A variant of the CS theory known as the BF theory captures the mutual statistics among different species of anyons, such as those of electric and magnetic quasiparticles in toric code (TC) or of vortex and quasiparticle in superfluids~\cite{hansson04}. In both CS and BF theories, the phase factor picked up in the process of braiding one anyon around other anyons is proportional to the number of anyon charges enclosed and can be understood within the flux attachment picture in which one anyon sees the effective magnetic flux tied to other anyons and picks up the corresponding Aharonov-Bohm (AB) phase. 

Recently, sub-dimensional topological order embodied in various exactly solvable spin models has emerged at the frontier of quantum matter research~\cite{haah11,vijay16,pretko17,slagle18,pretko18,hermele19,you20b,kim21}. These sub-dimensional topological orders are characterized by excitations such as fractons or lineons with constrained mobilities. Various exotic field theories attempting to capture aspects of fracton physics are under vigorous pursuit~\cite{slagle17, pretko18-2, chen18, you20a, hughes20, seiberg20, seiberg21a, seiberg21b, you21}. It was also realized that applying the Higgsing procedure to a lattice gauge theory (LGT) yields an exactly solvable spin model while lowering U(1) gauge symmetry to $\mathbb{Z}_N$~\cite{hermele18b,barkeshli18,oh-kim22,kim22}. A well-known X-cube model of fractons~\cite{vijay16} is obtained by Higgsing the rank-2 LGT composed of gauge fields with only off-diagonal indices~\cite{slagle17}. On the other hand, Higgsing the symmetric rank-2 LGT with both diagonal and off-diagonal gauge fields on the two-dimensional square lattice yields the rank-2 toric code (R2TC)~\cite{hermele18b,barkeshli18,oh-kim22}. In contrast to the parent gauge theory hosting immobile fractons or lineons, the excitations in R2TC are mobile and can hop by $N$ lattice spacings in the previously forbidden directions due to the condensation of charge-$N$ excitations in the $\mathbb{Z}_N$ gauge theory. An interesting consequence of it is the dependence of the  ground state degeneracy on the linear system size $( L_x , L_y )$ mod $N$, studied in detail in recent works~\cite{oh-kim22,seiberg22,pace22,chamon22}.

In this paper, we focus on the other exotic property of R2TC, {\it i.e.} mutual braiding statistics among the quasiparticles that seems to break the conventional wisdom of Abelian anyon braiding as well as the flux attachment picture. In R2TC, the braiding phase was found to depend on the initial positions of the quasiparticles participating in the braiding~\cite{oh-kim22}. Recall that the conventional anyon braiding counts the total charge of anyons inside the loop, and the statistical phase is encoded in the BF field theory. We call this framework the {\it monopolar braiding statistics}. Here we develop a new field-theoretic formalism consistent with what we call the {\it dipolar braiding statistics} and develop the associated {\it dipolar BF theory} of braiding. The new field theory is constructed from the rank-2 gauge fields, as the conventional BF theory is made out of rank-1 gauge fields. The flux attachment picture is similarly modified to that of dipole attachment. 

We begin by giving a quick review of the R2TC model in Sec. \ref{sec:r2tc-review}. In Sec. \ref{sec:dipolar-braiding}, we define this anomalous position dependent statistics as the dipolar braiding statistics. Construction of effective field theory is done in two steps. First, in Sec. \ref{sec:quasi-dynamics}, we construct the effective Lagrangian for matter fields minimally coupled to the rank-2 gauge fields. Then, in Sec. \ref{sec:dipolar-bf-theory,} we construct the full dipolar BF theory and derive dipolar braiding statistics from it. Summary and discussion follows in Sec. \ref{sec:discussion}.

\section{Review of R2TC}
\label{sec:r2tc-review}
The R2TC model is obtained by Higgsing the rank-2 U(1) LGT in two dimensions employing a pair of canonically conjugate fields $A^{ab}_i$ and  $E^{ab}_i$ obeying $[A, E]=i$ for the same site and indices. There are three independent components $(A^{xx}_i , A^{xy}_i , A^{yy}_i )$ per site $i$ (similarly for $E$). Several mutually commuting operators called generators can be constructed in the rank-2 LGT as: 
\begin{align}
G^{x}_i = &  E_{i+\hat{x}}^{xx} - E_{i}^{xx} + E_{i+\hat{y}}^{xy}- E_i^{xy}  \equiv \rho_i^x \nn 
\sim & \partial_x E^{xx}_i + \partial_y E^{xy}_i , \nn
G^{y}_i = & E_{i+\hat{x}}^{xy} - E_{i}^{xy} + E_{i+\hat{y}}^{yy}- E_i^{yy} \equiv \rho_i^y \nn 
\sim & \partial_x E^{xy}_i + \partial_y E^{yy}_i , \nn 
B_i = & A^{xx}_{i+\hat{y}}+ A^{xx}_{i-\hat{y}} -2 A^{xx}_{i} + A^{yy}_{i+\hat{x}}+ A^{yy}_{i-\hat{x}} -2 A^{yy}_{i} \nn  
& ~~ -A^{xy}_i + A^{xy}_{i-\hat{x}} + A^{xy}_{i-\hat{y}} - A^{xy}_{i-\hat{x}-\hat{y}} \equiv \frac{2\pi}{N}\rho_i^m \nn 
\sim &  \partial_y^2 A^{xx}_i + \partial_x^2 A^{yy}_i - \partial_x \partial_y A^{xy}_i .
\label{eq:r2lgt}
\end{align}
The theory contains a vector of electric charges $(\rho_i^x , \rho_i^y)$ and a scalar magnetic charge $\rho_i^m$, tied to the gauge fields through the Gauss law constraints $G_i^a = \rho_i^a$ ($a=x,y)$ and $B_i \equiv (2\pi/N) \rho_i^m$ for some integer $N\ge 2$. They generate the gauge transformations $A^{ab}_i \rightarrow  U_A^\dag A_i^{ab} U_A$ and $E^{ab}_i \rightarrow U_E^\dag E_i^{ab} U_E$ which result in 
\begin{align}
A^{aa}_i &\rightarrow A^{aa}_i + f^a_i - f^a_{i-\hat{a}} ~~ (a=x,y)\nn 
A^{xy}_i &\rightarrow A^{xy}_i + f^y_i - f^y_{i-\hat{x}}  + f^x_i - f^x_{i-\hat{y}} \nn 
E^{xx}_i &\rightarrow E^{xx}_i + g_{i+\hat{y}} + g_{i-\hat{y}} -2 g_i  \nn 
E^{yy}_i &\rightarrow E^{yy}_i + g_{i+\hat{x}} + g_{i-\hat{x}} -2 g_i  \nn 
E^{xy}_i &\rightarrow E^{xy}_i + g_{i+\hat{x}} +  g_{i+\hat{y}} - g_i - g_{i+\hat{x} +\hat{y}} .  \label{eq:A-and-E-transform}
\end{align}
The two unitary operators are 
\begin{align} U_A = e^{  i \sum_i ( f_i^x G_i^x  + f_i^y G_i^y ) }  ,  ~~
U_E = e^{ i \sum_i g_i B_i }  .  \label{eq:A-and-E-Unitary} 
\end{align} 

Generalized Pauli operators are obtained from Higging the gauge fields as~\cite{hermele18b,barkeshli18}:
\begin{align}
&X_{0,i} = e^{i A_i^{xy}}, &  &Z_{0,i} = e^{ 2\pi i E_i^{xy}/N}, \nn
&X_{1,i} = e^{i A_i^{xx}}, &  &Z_{1,i} = e^{ 2\pi i E_i^{xx}/N}, \nn
&X_{2,i} = e^{i A_i^{yy}}, &  &Z_{2,i} = e^{ 2\pi i E_i^{yy}/N}.
\label{eq:r2-Higgsing}
\end{align}
They operate on the $N$-dimensional local Hilbert space defined by 
\begin{align}
X|g\rangle = |g+1 \rangle, ~ Z|g\rangle  = \omega^g |g\rangle, ~ ZX = \omega XZ ,
\label{eq:ZN-spin}
\end{align}
where $|g\rangle = |0\rangle,~ |1\rangle,~ \cdots,~ |N-1\rangle$ and $\omega = e^{2\pi i /N}$. Applying the Higgsing formulas in Eq. (\ref{eq:r2-Higgsing}) to the three mutually commuting generators in Eq. (\ref{eq:r2lgt}) results in three types of commuting spin operators: 
\begin{align}
a_i^x &\equiv e^{ (2\pi i/N) G^{x} _i} =  Z_{1,i}^{-1} Z_{1,i+\hat{x}} Z_{0,i} Z_{0,i-\hat{y}}^{-1},\nn
a_i^y &\equiv e^{ (2\pi i/N) G^{y}_i} =  Z_{2,i}^{-1}  Z_{2,i+\hat{x}} Z_{0,i} Z_{0,i-\hat{x}}^{-1},\nn
b_i  &\equiv e^{ i B_i }  = X_{0,i}^{-1} X_{0,i-\hat{x}} X_{0,i-\hat{y}} X_{0,i-\hat{x}-\hat{y}}^{-1} \nn
&\times  X_{2,i-\hat{x}} X_{2,i}^{-2} X_{2,i+\hat{x}} 
X_{1,i-\hat{y}} X_{1,i}^{-2} X_{1,i+\hat{y}} . 
\label{eq:vector_Zn-operators}
\end{align}
The R2TC Hamiltonian is constructed as
\begin{align}
\mathbb{A}_i^x & = \frac{1}{N} \sum_{j =0}^{N-1} (a_i^x )^j ,    &\mathbb{A}_i^y & = \frac{1}{N} \sum_{j =0}^{N-1} (a_i^y )^j , \nn
\mathbb{B}_{i} & = \frac{1}{N} \sum_{j =0}^{N-1} (b_i )^j ,    &H & = -\sum_i (\mathbb{A}^x_i + \mathbb{A}^y_i + \mathbb{B}_i ) . 
\label{eq:r2tc}
\end{align}
This model has the ground states characterized by $\mathbb{A}_i^x |{\rm GS}\rangle = \mathbb{A}_i^y |{\rm GS} \rangle = \mathbb{B}_i |{\rm GS} \rangle = |{\rm GS} \rangle$. An excited state is obtained when one of the stabilizers $a_i^x$, $a_i^y$, or $b_i$ takes on the eigenvalue $\omega^j$ with nonzero integer $j$. We use $e_x$, $e_y$, and $m$ to denote the two electric and one magnetic charges associated with $a_i^x$, $a_i^y$, and $b_i$ excitations, respectively. The $e_x$ ($e_y$) monopole can move freely in the $x$-($y$-) direction, but can hop only by $N$ lattice spacings in the $y$- ($x$-) direction; the magnetic monopole can only hop by $N$ spacings in both directions~\cite{oh-kim22}.

\section{Dipolar Braiding Statistics}
\label{sec:dipolar-braiding}
It was shown~\cite{oh-kim22} that non-trivial braiding statistics arises between $(e_x ,m)$ and $(e_y , m)$ monopoles\footnote{We often refer to $e_x, e_y, m$ quasiparticle excitations as `monopoles' in order to distinguish them from dipole excitations that are composites of the monopoles.}  with phase factors given by
\begin{align}
e^{i  \phi_{e_x, m} } = \omega^{ (y_{x0} - y_{m0} )  }, ~~
e^{i \phi_{e_y, m} }  = \omega^{ (x_{m0} - x_{y0} ) } .  
\label{eq:braiding-statistics}
\end{align}
The $e_x$, $e_y$, and $m$ monopoles are located at 
\begin{align} {\bm r}_{x0} = (x_{x0} , y_{x0} ), ~ {\bm r}_{y0} = (x_{y0} , y_{y0}), ~ {\bm r}_m = (x_{m0} , y_{m0}) \end{align} 
at the start of the braiding, respectively. 

The following integral representations of the phase factors can be given for the braiding of $e_x$ and $e_y$ particle around $m$: 
\begin{align}
\phi_{e_x, m} & =   \frac{2 \pi}{N} \int d^2 {\bm r} ~ (y_{x0} - y ) \rho^m ({\bm r})  ,\nn
\phi_{e_y, m} & = \frac{2 \pi}{N}  \int d^2 {\bm r} ~ (x-x_{y0} ) \rho^m ({\bm r} )  .
\label{eq:phixphiy} 
\end{align}
The magnetic monopole density inside the contour is $\rho^m ({\bm r})$. The phase factors in Eq. (\ref{eq:braiding-statistics}) is recovered for a single $m$ monopole $\rho^m ({\bm r} ) = \delta^2 ({\bm r} - {\bm r}_{m0} )$. The initial position of the $e_x$ or $e_y$ particle that does the braiding enters explicitly in the above integrand. The familiar anyonic statistics is obtained instead from integrating the monopole density; in contrast Eq. (\ref{eq:phixphiy}) can be thought to integrate the {\it dipole density} of the enclosed particles. 

The phases $\phi_{m, e_x}$ and $\phi_{m, e_y}$ for braiding the $m$ monopole around $e_x$ and $e_y$ monopole, respectively, have similar integral expressions: 
\begin{align}
\phi_{m, e_x} =&  \frac{2\pi}{N} \int d^2 {\bm r}  (y-y_{m0} ) \rho^{x} ({\bm r}), \nn 
\phi_{m, e_y} =& \frac{2\pi}{N} \int d^2 {\bm r}  (x_{m0}-x ) \rho^{y} ({\bm r}) . 
\label{eq:phi-pxpy-braid}
\end{align}
The densities of $e_x$ and $e_y$ are  $\rho^{x} ({\bm r})$ and $\rho^{y} ({\bm r})$, respectively. 

A lattice-model interpretation of the integral formulas in Eqs. (\ref{eq:phixphiy}) and (\ref{eq:phi-pxpy-braid}) can be given. Discretizing the phase integrals in Eqs. (\ref{eq:phixphiy}) and (\ref{eq:phi-pxpy-braid}) and invoking the Higgsing formula in Eq. (\ref{eq:vector_Zn-operators}), 
\begin{align}
e^{i \phi_{e_x , m }} & \rightarrow W_{e_x , m } = \exp \left( i \frac{2\pi}{N} \sum_{i \in {\cal A}} (y_{x0} - y_i ) \rho_i^m  \right) \nn 
& ~~~~~~~~~~ =  \prod_{i \in \mathcal{A}} \left( b_i\right)^{y_{x0} -y_i } \nn 
e^{i \phi_{e_y , m }} & \rightarrow W_{e_y , m } = \exp \left( i \frac{2\pi}{N} \sum_{i \in {\cal A}} (x_i - x_{y0} ) \rho_i^m  \right) \nn 
& ~~~~~~~~~~ =  \prod_{i \in \mathcal{A}} \left( b_i\right)^{x_i - x_{y0} } \nn  
e^{i \phi_{m, e_x }} & \rightarrow W_{m,e_x} = \prod_{i \in {\cal A} } \left(a_{i}^x \right)^{y_i - y_{m0}}, \nn 
e^{i \phi_{m, e_y }} & \rightarrow W_{m,e_y} = \prod_{i \in {\cal A} } \left(a_{i}^y \right)^{x_{m0}-x_i}.
\label{eq:weg-loop-q}
\end{align}
Each operator $W$ is given as a product of various stabilizers over the area $\prod_{i \in \mathcal{A}}$ inside the braiding path. When acting on the ground state this yields 1, but not so if some monopoles reside in the area. Inserting explicit expressions for $b_{i}$ from Eq. (\ref{eq:vector_Zn-operators}) into $W_{e_x, m}$ and $W_{e_y, m}$ shows that all the operators in the interior of ${\cal A}$ cancel out, leaving only the product of $X$ operators along the boundary hence qualifying them as Wegner-Wilson operators of sorts. 

In the case of a rectangular boundary ${\cal A} = [x_0, x_1] \times [y_0 , y_1 ]$ one finds 
\begin{widetext} 
\begin{align}
W_{e_x , m} & = T_x (x_0\rightarrow x_1|y_0) T_x (y_0\rightarrow y_1|x_1 ) \left[T_x (x_0\rightarrow x_1| y_1)\right]^{-1} \left[T_x (y_0\rightarrow y_1|x_0)\right]^{-1}  \nn 
W_{e_y , m} & = T_y (x_0\rightarrow x_1|y_0) T_y (y_0\rightarrow y_1|x_1 ) \left[T_y (x_0\rightarrow x_1| y_1)\right]^{-1} \left[T_y (y_0\rightarrow y_1|x_0)\right]^{-1}
\label{eq:phix-decompose}
\end{align}
where the $e_x , e_y$ translation operators are defined as 
\begin{align}
& T_x (x_1\!\rightarrow\! x_2|y) \!=\! \prod_{x=x_1}^{x_2-1} \left( X_{1,i} \right)^{-1} \left( X_{1,i}X_{1,i \!-\! \hat{y} }^{-1}\right) ^{y\!-\!y_{x0}} \!=\! \exp \left( i \sum_{x = x_1}^{x_2-1}  \left[- A^{xx}(x,y)  \!+\! (y \!-\! y_{x0})( A^{xx}(x,y) \!-\! A^{xx}(x,y\!-\!1) )  \right] \right) , \nn
& T_x (y_1\!\rightarrow\! y_2|x) 
\!=\! 
\prod_{y=y_1}^{y_2-1} \!\left( X_{0,i} X_{0,i \!-\!\hat{y}}^{-1} X_{2,i}^{-1} X_{2,i\!-\!\hat{x}} \right)^{y\!-\!y_{x0}} 
\!= \!
\exp\! \left( \! i  \sum_{y \!=\! y_1}^{y_2-1} \! \left(y\!-\!y_{x0}\right)\!\left[ A^{xy}(x,\!y) \! - \!A^{xy}(x,\!y\!-\!1) \!-\! A^{yy}(x,\!y) \!+\! A^{yy}(x\!-\!1,\!y)  \right]\! \right) , \nn 
& T_y (x_1\!\rightarrow\! x_2|y) 
\!=\! 
\prod_{x=x_1}^{x_2-1}\!\left( X_{0,i} X_{0,i \!-\!\hat{x}}^{-1} 
X_{1,i}^{-1} X_{1,i\!-\!\hat{y}} \right)^{x\!-\!x_{y0}}
\!=\! 
\exp\! \left( \! i  \sum_{x \!=\! x_1}^{x_2-1} \! \left(x\!-\!x_{y0}\right)\!\left[ A^{xy}(x,\!y) \! - \!A^{xy}(x\!-\!1,\!y) \!-\! A^{xx}(x,\!y) \!+\! A^{xx}(x,\!y\!-\!1)  \right]\! \right) , \nn
& T_y (y_1\!\rightarrow\! y_2|x) \!=\!
\prod_{y=y_1}^{y_2-1} \left( X_{2,i} \right)^{-1} \left( X_{2,i}X_{2,i \!-\! \hat{x} }^{-1}\right) ^{x-x_{y0}}
\!= \! 
\exp \left( i \sum_{y= y_1}^{y_2-1}  \left[- A^{yy}(x,y)  \!+\! (x \!-\! x_{y0})( A^{yy}(x,y) \!-\! A^{yy}(x\!-\!1,y) )  \right] \right).
\label{eq:px-translate}
\end{align}
\end{widetext}
The $T_x (x_1\!\rightarrow\! x_2|y)$ translates the $e_x$ quasiparticle from $(x_1 , y)$ to $(x_2 , y)$. On the other hand, $T_x (y_1\!\rightarrow\! y_2|x)$ translates $e_x$ from $(x, y_1)$ to $(x, y_2)$ only if $y_1 - y_{x0} \md N = 0$ and $y_2 - y_{x0} \md N = 0$~\cite{oh-kim22}. Recall that the $e_x$ monopole is able to hop only by $N$ lattice sites in the $y$-direction. Nevertheless, one can define the operator $T_x (y_1\!\rightarrow\! y_2|x)$ for any pair of coordinates $y_1 , y_2$, under a new interpretation: for general $y_1, y_2$, the translation of $e_x$ embodies the simultaneous creation and motion of some auxiliary dipoles that are necessary to maintain the overall dipole moment conservation mod $N$. Similar statements apply to $T_y (x_1 \rightarrow x_2 | y)$. Details can be found in Appendix \ref{sec:emergent-gauge-derivation}.

On taking continuum limits of Eq. (\ref{eq:px-translate}), the braiding phases become line integrals 
\begin{align}
\phi_{e_x, m} & = \oint d {\bm r} \cdot {\bm a}_x , &  \phi_{e_y , m}   & = \oint  d{\bm r}\cdot {\bm a}_y
\end{align}
with a pair of {\it emergent} vector potentials defined by  
\begin{align}
{\bm a}_x =& \bigl( \!-\!A^{xx} \!+\! \left(y\!-\!y_{x0} \right)\! \partial_{y} A^{xx},
\! (y\!-\!y_{x0} ) \! \left( \partial_y A^{xy}\! -\! \partial_x A^{yy}  \right) \bigr), \nn 
{\bm a}_y  =& 
\bigl( \!(x\!-\!x_{y0} ) \! \left(  \partial_x A^{xy} \! -\! \partial_y A^{xx}  \right) ,
\!  -\! A^{yy} \!+\! \left(x\!-\!x_{y0} \right)\! \partial_{x} A^{yy} \bigr).
\label{eq:emergent-a}
\end{align}
The meaning of ${\bm a}_x$ and ${\bm a}_y$ is clarified on taking their curl,
\begin{align}
{\bm \nabla} \times {\bm a}_x =&   \frac{2\pi}{N} (y_{x0} -y ) \rho^m \nn 
{\bm \nabla} \times {\bm a}_y =&  \frac{2\pi}{N} (x-x_{y0} ) \rho^m ,
\label{eq:emergent-rho-q}
\end{align}
where we used the constraint 
\begin{align} \partial_x^2 A^{yy} + \partial_y^2 A^{xx} - \partial_x \partial_y A^{xy} =  (2\pi/N) \rho^m. \nonumber \end{align} 
In an ordinary flux attachment scenario, the curl of the emergent vector potential equals the charge density. Here it gives either the $x$ or the $y$ component of the dipole density. The initial positions of the $e_x$ and $e_y$ monopoles appear explicitly in the formulas as a result. 

Next, we turn to the braiding operators $W_{m, e_x}$ and $W_{m, e_y}$. It turns out $W_{e_x, m}$ and $W_{e_y , m}$ individually do not have the nice cancellation of bulk terms but their product $W_{m,e} \equiv W_{m,e_x}W_{m,e_y}$ does, and become a product of boundary operators
\begin{widetext}
\begin{align}
W_{m,e} & = T_m (x_0 \rightarrow x_1 | y_0) T_m (y_0 \rightarrow y_1 | x_1 ) \left[T_m (x_0 \rightarrow x_1 | y_1) \right]^{-1} \left[ T_m (y_0 \rightarrow y_1 | x_0 ) \right]^{-1} ,
\label{eq:wqpxpy-prod}\end{align}
with 
\begin{align}
&T_m (x_1 \rightarrow x_2 | y) = \prod_{x=x_1\!+\!1}^{x_2}  \left(Z_{2,i} \right)^{x \!-\! x_{m0}} \left( Z_{0,i-\hat{x}}\right)^{y_{m0}\!-\!y }
\!=\!\exp \left[i \frac{2\pi}{N} \sum_{x=x_1\!+\!1}^{x_2} \bigl((x\!-\!x_{m0} ) E^{yy}(x,y) \!-\! (y\!-\!y_{m0} )E^{xy}(x\!-\!1,y)   \bigr) \right] , \nn 
& T_m (y_1 \rightarrow y_2 | x) = \prod_{y=y_1\!+\!1}^{y_2}\left( Z_{1,i} \right)^{y \!-\!y_{m0} }  \! \left(Z_{0,i-\hat{y}} \right)^{x_{m0}\!-\!x } 
\!=\! \exp\! \left[ i  \frac{2\pi}{N}\!  \sum_{y=y_1\!+\!1}^{y_2} \! \bigl((y\!-\!y_{m0} ) E^{xx}(x,y) \!-\! (x\!-\!x_{m0} )E^{xy}(x,y\!-\!1) \bigr)\! \right].
\label{eq:m-translate}
\end{align}
\end{widetext}
The overall braiding phase $\phi_{m , e} = \phi_{m , e_x } + \phi_{m , e_y }$ for $m$ around both $e_x$ and $e_y$ can be written  $\phi_{m, e} = \oint d{\bm r} \cdot {\bm a}_m$,
where 
\begin{align}
{\bm a}_m \!=\! \frac{2\pi}{N}
& \bigl( (x\!-\!x_{m0} ) E^{yy}\!-\!(y\! -\!y_{m0} )E^{xy}  , \nn 
& ~~~~~~~~~~ (y\!-\!y_{m0} ) E^{xx} \!-\!(x\!-\!x_{m0} )E^{xy} \bigr) , \nn 
{\bm \nabla} \times {\bm a}_m  & = \frac{2\pi}{N} \left[  (y-y_{m0} ) \rho^{x} -(x-x_{m0} )\rho^{y}  \right] .
\label{eq:emergent-rho-p}
\end{align}
The constraints 
\begin{align} \partial_x E^{xx} \!+\! \partial_y E^{xy} = \rho^{x}, ~~
\partial_x E^{xy} \!+\! \partial_y E^{yy} = \rho^{y}  \nonumber \end{align} were used in the last line. 

To sum up, a {\it dipolar AB phase} formulation of the dipolar braiding statistics is possible in terms of some emergent vector potential ${\bm a}_x$, ${\bm a}_y$ and ${\bm a}_m$, which are functions of the underlying rank-2 fields $A^{ab}, E^{ab}$ as well as the initial locations of the quasiparticles. For completeness we mention that under the gauge transformation 
\begin{align}
A^{xx}\rightarrow & A^{xx} + \partial_x f_x, 
& E^{xx}\rightarrow & E^{xx} + \partial_y^2 g,
\nn 
A^{xy}\rightarrow & A^{xy} + \partial_x f_y \!+\! \partial_y f_x,  &
E^{xy}\rightarrow & E^{xy}  - \partial_x \partial_y g,
\nn 
A^{yy}\rightarrow & A^{yy} +  \partial_y f_y,  &
E^{yy}\rightarrow & E^{yy} +  \partial_x^2 g, \nonumber
\end{align}
of the rank-2 fields, 
the  emergent vector potentials transform as
\begin{align}
    {\bm a}_x \rightarrow  {\bm a}_x + {\bm \nabla}  F_x , 
    {\bm a}_y \rightarrow  {\bm a}_y +  {\bm \nabla} F_y ,  
    {\bm a}_m \rightarrow {\bm a}_m + {\bm \nabla} G , \nonumber 
\end{align}
where
\begin{align}
( F_x , F_y ) &= \bigl( -f_x  + (y-y_{x0} )\partial_{y}f_x , f^y - (x-x_{y0} ) \partial_{x}f_y \bigr)  \nn 
G & = (x - x_{m0}) \partial_x g + (y-y_{m0}) \partial_y g - g. \nonumber 
\end{align}

\section{Matter Fields with Minimal Coupling}
\label{sec:quasi-dynamics}

There are three matter fields in the R2TC, represented by field operators $(\psi_x, \psi_y , \psi_m )$, which are related to the monopole density by $\rho^a = \psi_a^\dag \psi_a$ $(a=x,y,m)$. Appropriate field theory for these can be constructed by exploiting the gauge transformation properties in Eqs. (\ref{eq:A-and-E-transform}) and (\ref{eq:A-and-E-Unitary}) and identifying proper covariant derivatives of the matter fields. 

Taking the continuum limit of the unitary operators in Eq. (\ref{eq:A-and-E-Unitary}) and invoking the constraints $G_i^a = \rho_i^a$ and $B_i = (2\pi/N) \rho_i^m$ from Eq. (\ref{eq:r2lgt}), one can show that the matter fields $\psi_a$ $(a=x,y,m)$ and the gauge fields $(A^{ab}, E^{ab})$ transform as 
\begin{align} 
(\psi_x, \psi_y , \psi_m ) & \rightarrow (e^{if_x} \psi_x, e^{if_y} \psi_y , e^{i\frac{2\pi}{N}g} \psi_m )  \nn 
A^{ab} & \rightarrow A^{ab}\!+\!( \partial_x f_x, \partial_x f_y \!+\! \partial_y f_x , \partial_y f_y ) \nn 
E^{ab} & \rightarrow E^{ab}\!+\! (\partial_y^2 g, - \partial_x \partial_y g, \partial_x^2 g ) . 
\label{eq:gauge-transf-Gp}
\end{align}
Identifying the covariant derivatives and constructing a gauge-invariant Lagrangian ia a straightforward exercise~\cite{kim22}. For $e_x$ $(\psi_x)$ and $e_y$ ($\psi_y$) particles they are 
\begin{align}
& D_x \psi_{x} \equiv (\partial_x - i A^{xx} )\psi_{x},\nn 
& D_y \psi_{y} \equiv (\partial_y - i A^{yy} ) \psi_{y} , \nn 
& D_{xy} \psi_{xy} \equiv \psi_{x}\partial_x \psi_{y} + \psi_{y} \partial_y \psi_{x} - i A^{xy} \psi_{x} \psi_{y} ,  \end{align}
and
\begin{align}
& L_e = \sum_{a=x,y} \psi_{a}^\dag \partial_t \psi_{a} \nn 
& - \frac{1}{2} \left( \alpha_e \left| D_x \psi_{x}\right|^2 + \alpha_e \left|D_y \psi_{y} \right|^2 + \beta_e \left| D_{xy}\psi_{xy} \right|^2  \right) .  \label{eq:Le}
\end{align}
Similarly for the $m$ particles,
\begin{align} 
& D_{xx} \psi_m \equiv  \psi_m \partial_x^2 \psi_m - (\partial_x \psi_m )^2 - i\frac{2\pi}{N} E^{yy} \psi_m^2 , \nn 
& D_{yy} \psi_m \equiv  \psi_m \partial_y^2 \psi_m - (\partial_y \psi_m )^2 - i\frac{2\pi}{N} E^{xx} \psi_m^2 , \nn 
& D_{xy} \psi_m  \equiv  \psi_m \partial_x \partial_y \psi_m - (\partial_x \psi_m )(\partial_y \psi_m ) + i\frac{2\pi}{N} E^{xy} \psi_m^2 \end{align}
and
\begin{align}
& L_m = i \psi_m^\dag \partial_t \psi_m \nn 
& - \frac{1}{4} \left( \alpha_m \left| D_{xx}\psi_m \right|^2  + \alpha_m \left| D_{yy}\psi_m \right|^2+ \beta_m \left| D_{xy}\psi_m \right|^2  \right) .  \label{eq:Lm} 
\end{align}
define the appropriate matter field theory. Various constants ($\alpha$'s $\beta$'s) appear, which do not change the generic features of the field theory. 

Continuity equations for the quasiparticles particles follow as 
\begin{align}
\partial_t {\rho}^{x}+ \partial_x J_e^{xx} + \partial_y J_e^{xy} & = 0, \nn 
\partial_t {\rho}^{y}+ \partial_x J_e^{xy} + \partial_y J_e^{yy} & = 0, \nn 
\partial_t {\rho}^m + \partial_x^2 J_m^{xx} +\partial_x \partial_y J_m^{xy} + \partial_{y}^2 J_m^{yy} &  = 0,
\label{eq:continuity}
\end{align}
where
\begin{align}
J_e^{aa}  \equiv & - \frac{i}{2} \alpha_e \Big[ \psi_{a}^\dag \partial_a \psi_{a} - \left( \partial_a \psi_{a}^\dag \right) \psi_{a}  \Big], ~~ (a=x,y) \nn
J_e^{xy}  \equiv & - \frac{i}{2} \beta_e \psi_{x}^\dag \psi_{y}^\dag \left( \psi_{x} \partial_x \psi_{y} + \psi_{y}\partial_y \psi_{x}\right) + h.c.  =  J_e^{yx}, \nn 
J_m^{aa} = & \!-\!\frac{i}{4} \alpha_m \! \left(\psi_m^\dag \partial_a^2 \psi_m^\dag \!-\!\partial_a \psi_m^\dag \partial_a \psi_m^\dag  \right) \psi_m^2 \!+\! h.c. ~ (a\!=\!x,y) \nn 
J_m^{xy} = & \!-\!\frac{i}{4} \beta_m \!\left(\psi_m^\dag \partial_x \partial_y \psi_m^\dag \!-\!\partial_x \psi_m^\dag \partial_y\psi_m^\dag  \right) \psi_m^2 + h.c.  \!=\! J_m^{yx} .
\end{align}
Both the covariant derivatives and the conservation laws depart significantly from those of particles in the rank-1 U(1) gauge fields. 

Several conserved quantities can be identified from the continuity equations. Writing the three monpole charges $Q^a = \int d^2 {\bm r} \rho^a$ $(a=x,y,m)$, and three dipole charges
\begin{align}
{\bm \mu}^m  & = \int d^2 {\bm r} {\bm r}  \rho^m , \nn 
{\mu}^e & = \int d^2 {\bm r} (x \rho^y - y \rho^x ) , 
\end{align} 
one can show their time derivatives vanish identicallly under the appropriate boundary conditions at infinity. The existence of six conserved quantities in the theory is closely tied to the existence of (up to) six independent holonomies and (up to) $N^6$ ground state degeneracy in the R2TC\footnote{There is an intricate relation between the six conserved quantities mentioned here and the holonomies that generate the ground state degeneracy. They will be discussed in an upcoming article~\cite{oh-forthcoming}}.

Finally, the covariant derivatives and the accompanying Lagrangians we constructed can be cast in the lattice mode. Writing ($\psi_x, \psi_y, \psi_m  \rightarrow X_i, Y_i , M_i$), we obtain 
\begin{align} | D_x \psi_{x} |^2 & \rightarrow X^\dag_i X_{i-\hat{x}} e^{iA^{xx}_i} + h.c. \nn 
| D_y \psi_{y} |^2 & \rightarrow Y^\dag_i Y_{i-\hat{y}} e^{iA^{yy}_i } + h.c. \nn
|D_{xy} \psi_{xy} |^2 & \rightarrow Y^\dag_i Y_{i-\hat{x}} X^\dag_i X_{i-\hat{y}} e^{iA^{xy}_i} + h.c.  \nn 
|D_{xx} \psi_m |^2 & \rightarrow  M_i^2 M^\dag_{i+\hat{x}} M^\dag_{i-\hat{x}} e^{iE^{xx}_i} + h.c.  \nn 
|D_{yy} \psi_m |^2 & \rightarrow M_i^2 M^\dag_{i+\hat{y}} M^\dag_{i-\hat{y}} e^{iE^{yy}_i} + h.c. \nn 
|D_{xy} \psi_m |^2 & \rightarrow M_i M_{i+\hat{x}+\hat{y}} M^\dag_{i+\hat{x}} M^\dag_{i+\hat{y}} e^{iE^{xy}_i} + h.c. \end{align} 
Similar expressions have been suggested in the tight-binding model of higher-order topological insulator~\cite{you21}. 

\section{Dipolar BF theory}
\label{sec:dipolar-bf-theory}
Comparing the temporal derivatives of the constraints in Eq. (\ref{eq:r2lgt}) with the continuity equation in Eq. (\ref{eq:continuity}) allows the identification of several identities: 
\begin{align} J_e^{ab} = - \partial_t {E}^{ab}, \frac{2 \pi}{N} J_m^{xx} = -\partial_t {A}^{yy}, \nn 
\frac{2 \pi}{N} J_m^{xy} = \partial_t {A}^{xy}, \frac{2 \pi}{N} J_m^{yy} = -\partial_t {A}^{xx}. \end{align} 
The Lagrangian that encodes all these constraints as well as the commutation $[A^{ab} ({\bm r}) ,E^{ab} ({\bm r}' ) ]=i \delta^2 ({\bm r} - {\bm r}' )$ is
\begin{widetext}
\begin{align}
{\cal L}_{\rm dBF} & =  A^{0x}\left( \partial_x E^{xx} \!+\! \partial_y E^{xy} -\rho^x \right) + A^{0y} \left( \partial_x E^{xy} \!+\! \partial_y E^{yy} -  \rho^y \right) + E^0\left( \partial_x^2 A^{yy} \!+\! \partial_y^2 A^{xx} \!-\! \partial_x \partial_y A^{xy}  - \frac{2 \pi}{N} \rho^m \right) \nn 
& + E^{xx}\left(\partial_t {A}^{xx} \!+\! \frac{2 \pi}{N} J_m^{yy}\right) 
  + E^{xy}\left(\partial_t {A}^{xy} \!-\! \frac{2 \pi}{N} J_m^{xy}  \right) 
  + E^{yy}\left(\partial_t {A}^{yy} \!+\! \frac{2 \pi}{N} J_m^{xx}\right) 
  - A^{xx} J_e^{xx} \!-\! A^{xy} J_e^{xy} \!-\! A^{yy} J_e^{yy} .
\label{eq:eff-lagr}
\end{align}
\end{widetext}
This action is the {\it dipolar BF theory} (dBF) in contrast to the BF theory of the rank-1 TC. The Lagrange multipliers transform as 
\begin{align}
    A^{0a} &\!\rightarrow\! A^{0a} \!+\! \partial_t f^a , &E^0 \!\rightarrow\! E^0 \!+\! \partial_t g.
\end{align}
The conservation laws of Eq. (\ref{eq:continuity}) are recovered from the gauge invariance of the action. 

The dipolar AB phase factors follow straightforwardly from the dBF action, but before doing so one must first address the important conceptual question: how can one adiabatically move a particle when its motion is constrained? For instance, the $m$ particle cannot move at all in the limit $N\rightarrow \infty$ due to the dipole conservation, and $e_x$ ($e_y$) can only move smoothly in the $x$ ($y$) direction. Nevertheless, by making use of the translation operators defined in Eqs. (\ref{eq:px-translate}) and (\ref{eq:m-translate}), it seems as though the monopoles can be freely moved around. Upon careful analysis of the lattice translation operators, we realize that such motion becomes possible at the expense of creating an auxiliary dipole that compensates for the changes in the dipole moment during the monopole motion, as explicitly demonstrated in Appendix \ref{sec:aux-dipole}. The composite particle of a monopole and an auxiliary dipole carries a net dipole moment of zero, and are free from constraint. 

We show how to derive the dipolar braiding phase explicitly in the the continuum theory. As an $m$ particle initially at ${\bm r}_{m0} = (x_{m0} , y_{m0} )$ is moved to ${\bm r}_m (t) = (x_m (t) , y_m (t) )$, the monopole and auxiliary dipole composite has the density 
\begin{align}
\rho^m ({\bm r}) =  [ 1 +( {\bm r}_m (t) - {\bm r}_{m0} ) \cdot {\bm \nabla} ]  \delta^2({\bm r}-{\bm r}_m (t) ) .
\end{align}
The second part reflects the dipole contribution. Details of the derivation can be found in Appendix \ref{sec:aux-dipole}. The total dipole moment ${\bm \mu}_m = \int d^2 {\bm r} {\bm r} \rho^m ({\bm r})$ is indeed conserved. When the position ${\bm r}_m$ varies over time, the time derivative of the total density $\rho^{m}$ becomes 
\begin{widetext}
\begin{align}
\partial_t \rho^{m}({\bm r}) =& -  \dot{x}_m \left[(x_m -x_{m0}) \partial_x^2 \delta^2({\bm r}-{\bm r}_m) + (y_m - y_{m0}) \partial_x \partial_y \delta^2({\bm r}-{\bm r}_m )  \right] \nn 
&- \dot{y}_m \left[(x_m -x_{m0}) \partial_x  \partial_y \delta^2({\bm r}-{\bm r}_m ) + (y_m -y_{m0})  \partial_y^2 \delta^2({\bm r}-{\bm r}_m ) \right].
\label{eq:rho_m_t_dev}
\end{align}
\end{widetext}
Invoking the continuity equation for $\rho^m$ in Eq. (\ref{eq:continuity}), one can deduce the accompanying current density
\begin{align}
J_m^{xx} & = \dot{x}_m (x_m-x_{m0}) \delta^2({\bm r} - {\bm r}_m), \nn  
J_m^{xy} & = \left(\dot{x}_m (x_m-x_{m0}) + \dot{y}_m (y_m-y_{m0}) \right)\delta^2({\bm r} - {\bm r}_m), \nn 
J_m^{yy} &= \dot{y}_m (y_m-y_{m0}) \delta^2({\bm r} - {\bm r}_m).
\label{eq:current-m}
\end{align}
Finally, inserting the obtained current density into the dBF action gives 
\begin{align}
&\frac{2\pi}{N}\int dt \int d^2 {\bm r} \left[E^{xx}J_m^{yy} - E^{xy} J_m^{xy}+E^{yy}J_m^{xx} \right] \nn 
& = \oint d{\bm r}_m \cdot {\bm a}_m ({\bm r_m} ) ,
\end{align}
precisely equal to the statistical phase $\phi_{m, e}$ from braiding $m$ around $e_x , e_y$ quasiparticles.

Likewise when the $e_x$  monopole is translated from ${\bm r}_{x0} = (x_{x0},y_{x0})$ to ${\bm r}_x (t) = (x_x (t) , y_x (t) )$, or the $e_y$ monopole from ${\bm r}_{y0} = (x_{y0},y_{y0})$ to ${\bm r}_y (t) = (x_y (t) ,y_y (t) )$, the net monopole and auxiliary dipole composite has the density
\begin{align}
\rho^x ({\bm r}) & = \delta^2({\bm r}\!-\!{\bm r}_x (t) ) + (y_x (t) \!-\!y_{x0}) \partial_y \delta^2({\bm r}-{\bm r}_x (t) ) , \nn 
\rho^y ({\bm r}) & = \delta^2({\bm r}\!-\!{\bm r}_y (t) ) + (x_y (t) \!-\!x_{y0}) \partial_x \delta^2({\bm r}\!-\!{\bm r}_y (t) ) . 
\end{align}
The net electric dipole moment $\mu_e = \int d^2 {\bm r} [ x \rho^y ({\bm r}) - y \rho^x ({\bm r}) ]$ is conserved during the motion. 

The current density accompanying the adiabatic motion of $e_x$ is 
\begin{align}
J^{xx}_e =& \dot{x}_x \left[ 1+(y_x-y_{x0}) \partial_y \right]  \delta^2({\bm r}-{\bm r}_x) , \nn 
J^{xy}_e =& \dot{y}_x (y_x -y_{x0}) \partial_y \delta^2({\bm r}-{\bm r}_x) , \nn 
J^{yy}_e =& -\dot{y}_x (y_x -y_{x0}) \partial_x \delta^2({\bm r}-{\bm r}_x) ,
\label{eq:current-ex}
\end{align}
while that of the $e_y$ is 
\begin{align}
J^{xx}_e =& -\dot{x}_y (x_y -x_{y0}) \partial_y \delta^2({\bm r}-{\bm r}_y), \nn 
J^{xy}_e =& \dot{x}_y (x_y -x_{y0}) \partial_x \delta^2({\bm r}-{\bm r}_y) , \nn 
J^{yy}_e =& \dot{y}_y \left[ 1+(x_y-x_{y0}) \partial_x \right]  \delta^2({\bm r}-{\bm r}_y) .
\label{eq:current-ey}
\end{align}
Inserting the current densities of the $e_x$ or $e_y$ monopoles in the dBF action results in
\begin{align}
-\int dt \int dxdy \left[ A^{xx} J_e^{xx} + A^{xy} J_e^{xy} + A^{yy} J_e^{yy} \right]
\end{align}
equal to either $\oint d{\bm r}_x \cdot {\bm a}_y({\bm r}_x) = \phi_{e_x , m}$ or 
$\oint d{\bm r}_y \cdot {\bm a}_y({\bm r}_y) = \phi_{e_y,m}$. The dBF action is thus consistent with the dipolar braiding phases obtained from the analysis of the lattice model. Importantly, both the monopole and the auxiliary dipole components in the density must be kept in order to derive the correct braiding statistics. 

\section{Discussion}
\label{sec:discussion}
Field-theoretic formulation for the dipolar braiding statistics first found in the R2TC model is developed. 
The statistical phases count the net dipole moment rather than the charge of the quasiparticles enclosed in the braiding. The adiabatic motion of a quasiparticle is possible when accompanied by the simultaneous motion of auxiliary dipole, which helps conserve the total dipole moment. The dipolar braiding statistics is derived by solving the equation of motion for quasiparticles. 

Various field theories of dipolar nature have been proposed in the past in three dimensions~\cite{you20a,you21}. Our paper proposes a dipolar field theory in two dimensions along with the accompanying dipolar statistics. While this work was under review, several papers analyzed the ground state degeneracy and the dipolar braiding statistics from a complementary perspective of anyon lattice~\cite{pace22,chamon22}, and proposed a Chern-Simons-type field (Ref. \cite{pace22}) or a combination of Chern-Simons and dipolar Chern-Simons theory descriptions (Ref. \cite{chamon22}). It remains to explore how the dipolar BF theory proposed in this work relates to these other theories. An important part of the specification of the topological field theory is that of the global structure under the so-called large gauge transformation. We have not completely worked out the global structure of the dipolar BF theory or the related ground state degeneracy count in this paper. These problems will be addressed in a forthcoming article~\cite{oh-forthcoming}. 

\acknowledgments 
Y.-T. O. was supported by National Research Foundation of Korea under Grant NRF-2014R1A6A1030732, NRF-2021R1A2C4001847, NRF-2020R1A4A3079707, and NRF-2022R1I1A1A01065149. 
JHH was supported by  NRF-2019R1A6A1A10073079. He also acknowledges financial support from EPIQS Moore theory centers at MIT and Harvard. J. K. was supported by the quantum computing technology development program of the National Research Foundation of Korea(NRF) funded by the Korean government (Ministry of Science and ICT(MSIT)). (No.2021M3H3A103657312) Additional support came from National Research Foundation of Korea under Grant NRF-2021M3E4A1038308 and NRF-2021M3H3A1038085. JHH acknowledges informative discussion with Ho Tat Lam, Salvatore Pace, and Yizhi You.

\appendix

\section{Emergent gauge fields from lattice consideration} 
\label{sec:emergent-gauge-derivation}

\begin{figure*}[tb]
\includegraphics[width=1\textwidth]{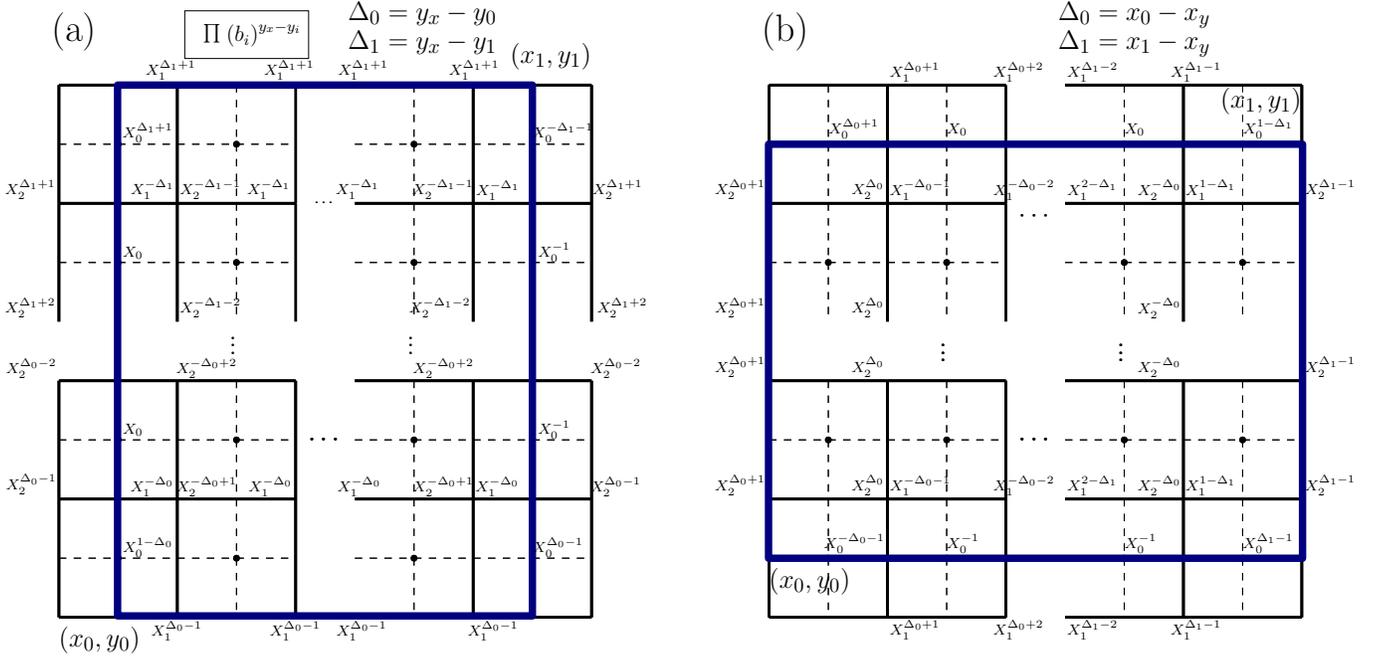}
\caption{
Schematic diagram for the operators (a) $W_{e_x,m}$ (c) and $W_{e_y,m}$ in Eq. (\ref{eq:weg-loop-e}). The boundary of WW operators are expressed by the navy lines.
}
\label{fig:wex-wey}
\end{figure*}

In this section, we elucidate the procedure by which the emergent gauge fields are obtained from the lattice consideration. Initially the discretized phase integrals were written as areal products $W_{e_x,m}$ and $W_{e_y,m}$:
\begin{align}
W_{e_x , m }=&  \prod_{i \in \cal{A}} \left( b_i\right)^{y_{x0} -y_i }, &
 W_{e_y , m } =&  \prod_{i \in \cal{A}} \left( b_i\right)^{x_i - x_{y0} }. \label{eq:weg-loop-e}
\end{align}
By substituting the definition of $b_i$ in terms of spin operators, we obtain equivalent, Wegner-Wilson (WW) loop expressions as a product of various $X$ operators along the boundary. 
We consider the case of a rectangular boundary ${\cal A} = [x_0, x_1] \times [y_0 , y_1 ]$ as depicted in Fig. \ref{fig:wex-wey}.
We can further decompose them into products of four line operators as
\begin{widetext}
\begin{align}
W_{e_x , m} & = T_x (x_0\rightarrow x_1|y_0) T_x (y_0\rightarrow y_1|x_1 )  \left[T_x (x_0\rightarrow x_1| y_1)\right]^{-1} \left[T_x (y_0\rightarrow y_1|x_0)\right]^{-1} , \nn 
W_{e_y , m} & = T_y (x_0\rightarrow x_1|y_0) T_y (y_0\rightarrow y_1|x_1 )  \left[T_y (x_0\rightarrow x_1| y_1)\right]^{-1} \left[T_y (y_0\rightarrow y_1|x_0)\right]^{-1} . 
\label{eq:weg-t-decompose}
\end{align}
\end{widetext}
For instance, $T_x (x_0 \rightarrow x_1 |y_0)$ involves the product of operators at $x=x_0$ through $x=x_1$ for a fixed $y=y_0$. Specifically, 
\begin{align}
& T_x (x_1\rightarrow x_2|y) \!=\! \prod_{x=x_1}^{x_2-1} \left( X_{1,i} \right)^{-1} \left( X_{1,i}X_{1,i \!-\! \hat{y} }^{-1}\right) ^{y-y_{x0}}, \nn
& T_x (y_1\rightarrow y_2|x) \!=\! \prod_{y=y_1}^{y_2-1} \left( X_{0,i} X_{0,i \!-\!\hat{y}}^{-1} X_{2,i}^{-1} X_{2,i\!-\!\hat{x}} \right)^{y\!-\!y_{x0}}, 
\label{eq:e-translate-2}
\end{align}
and,
\begin{align}
& T_y (x_1\rightarrow x_2|y) \!=\! \prod_{x=x_1}^{x_2-1} \left( X_{0,i} X_{0,i \!-\!\hat{x}}^{-1} X_{1,i}^{-1} X_{1,i\!-\!\hat{y}} \right)^{x\!-\!x_{y0}}, \nn
& T_y (y_1\rightarrow y_2|x) \!=\! \prod_{y=y_1}^{y_2-1} \left( X_{2,i} \right)^{-1} \left( X_{2,i}X_{2,i \!-\! \hat{x} }^{-1}\right) ^{x-x_{y0}}.
\label{eq:e-translate-1}
\end{align}
The shape of the boundary can be relaxed. For any type of boundary made by putting together several rectangles of arbitrary sizes, one still finds complete cancellation of operators except those at the boundary, which can be decomposed into the products of $T$ operators given in Eqs. (\ref{eq:e-translate-2}) and (\ref{eq:e-translate-1}).

\begin{figure*}[tb]
\includegraphics[width=0.8\textwidth]{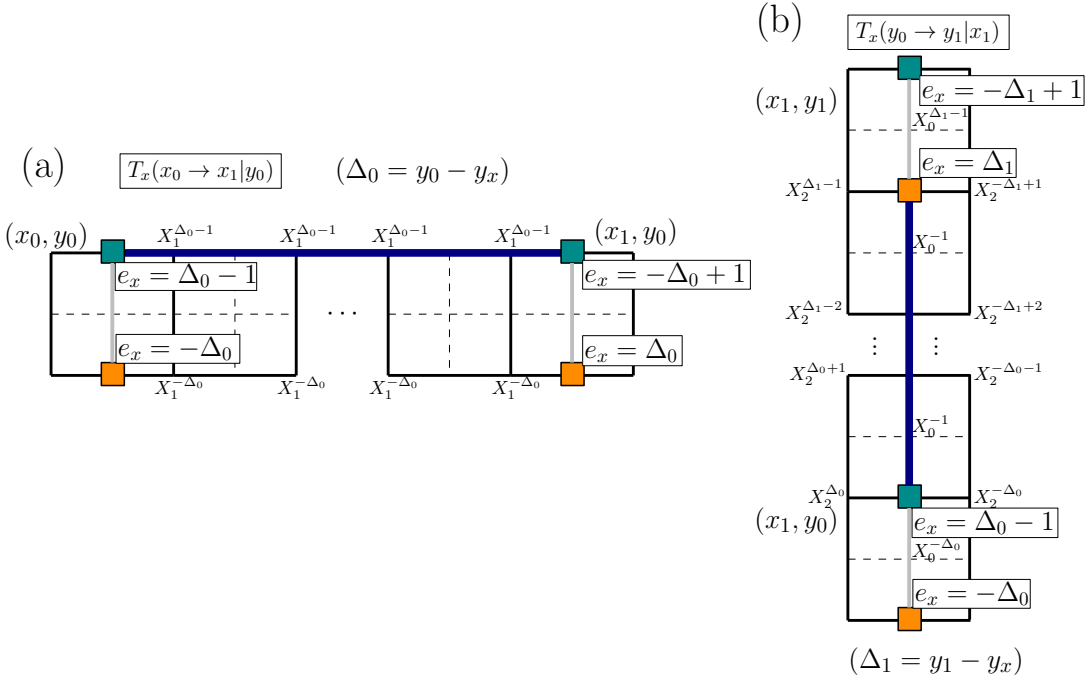}
\caption{Graphical illustration for (a) $T_x(x_0 \rightarrow x_1|y_0)$ and (b) $T_x(y_0 \rightarrow y_1|x_1)$. The navy line represents the domain of $T_x$. The cyan square represents $e_x$ monopole with charge $\pm 1$ together with other charges comprising the auxiliary dipoles. The orange square represents the other pair that make up the auxiliary dipole. The auxiliary dipoles are connected by the gray line. 
}
\label{fig:tx}
\end{figure*}

The first operator in $W_{e_x,m}$ is $T_x (x_0 \rightarrow x_1 | y_0 )$, which is illustrated in Fig. \ref{fig:tx}(a). It reduces to $\prod_{x=x_0}^{x_1-1} \left( X_{1,i} \right)^{-1}$ when $y_0 -y_{x0} \md N = 0$ and creates $e_x$ monopoles with charge $-1$ at $(x_0,y_0)$ and charge $+1$ at  $(x_1,y_0)$. In other words, the operator translates the $e_x$ monopole of charge +1 from $( x_0 , y_0) $ to $(x_1 , y_0)$. For $y_0-y_{x0} \md N \neq 0$, we need to take account of the action by the auxiliary factor $\left( X_{1,i}X_{1,i \!-\! \hat{y} }^{-1}\right) ^{y_0-y_{x0}}$ as well, which is the creation of a pair of $e_x$ dipoles. One of the dipoles created consists of two $e_x$ monopoles with charges $(y_0-y_{x0})$ at  $(x_0,y_0)$ and $-(y_0-y_{x0})$ at  $(x_0, y_0-1)$. The other dipole consists of two $e_x$ monopoles with charges $-(y_0-y_{x0})$ at  $(x_1,y_0)$ and $(y_0-y_{x0})$ at  $(x_1,y_0-1)$. 

We next look into the action of $T_x (y_0\rightarrow y_1| x_1)$ operator, which is illustrated in Fig. \ref{fig:tx}(b). First of all one can write $T_x (y_0\rightarrow y_1| x_1) = T_x (y_0 \rightarrow y_{x0} | x_1 ) T_x (y_{x0} \rightarrow y_1 | x_1 )$. When both $y_0-y_{x0} \md N = 0$ and $y_1-y_{x0} \md N =0$, one can think of it as a product of two translation operators of $e_x$, namely $T_x (y_0 \rightarrow y_{x0} | x_1 )$ and $T_x (y_{x0} \rightarrow y_1 | x_1 )$, moving $e_x$ from $(x_1 , y_0)$ to $(x_1 , y_{x0})$ and then subsequently from $(x_1 , y_{x0})$ to $(x_1, y_1)$. When $y_0-y_{x0} \md N \neq 0$ or $y_1-y_{x0} \md N \neq 0$, either the action of $T_x (y_0 \rightarrow y_{x0} | x_1 )$ or $T_x (y_{x0} \rightarrow y_1 | x_1 )$ is accompanied by the creation of an auxiliary dipole near $(x_1,y_0)$ or $(x_1,y_1)$, respectively. We can understand the action of $T_x$ and its inverse $T_x^{-1}$ as performing the translation of $e_x$ monopole while either creating or annihilating an auxiliary dipole. At the end of the $W_{e_x,m}$ operation, all the auxiliary dipoles disappear and the $e_x$ monopole has completed a loop.

By following the same procedure, one can interpret $T_y$ as the translation operator moving the $e_y$ monopole from one point to another and with or without the accompanying dipole depending on the modality of the coordinates with respect to $N$. The auxiliary dipoles are aligned in the $x$-direction and disappear at the completion of the loop.

\begin{figure*}[tb]
\includegraphics[width=0.8\textwidth]{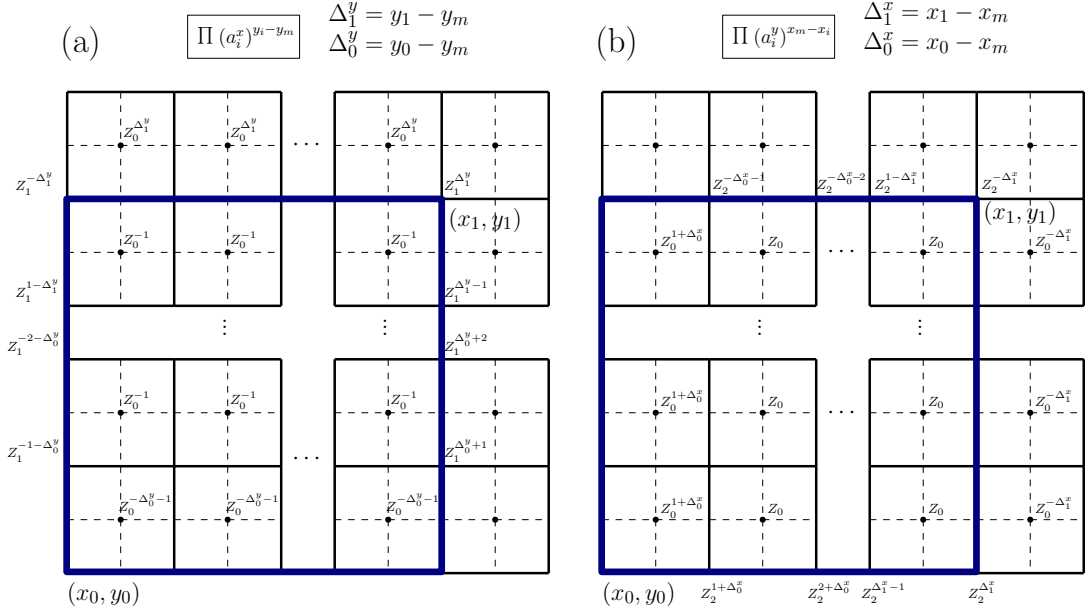}
\caption{
Graphical illustrations for the operators (a) $W_{m,e_x}$ and (b) $W_{m,e_y}$ in Eq. (\ref{eq:weg-loop-m}). The boundary of WW operators are expressed by the navy lines.
}
\label{fig:weg-loop-e}
\end{figure*}

The operators $W_{m,e_x}$ and $W_{m,e_y}$ are defined as
\begin{align}
W_{m,e_x} = & \prod_{i \in {\cal A} } \left(a_{i}^x \right)^{y_i - y_{m0}}, & 
W_{m,e_y} = & \prod_{i \in {\cal A} } \left(a_{i}^y \right)^{x_{m0} -x_i}.
\label{eq:weg-loop-m}
\end{align}

In the case of the rectangular boundary ${\cal A} = [x_0, x_1] \times [y_0 , y_1 ]$,  $W_{m,e_x}$ and $W_{m,e_y}$ are expressed in Fig. \ref{fig:weg-loop-e}. The presence of $Z_0$ operators in the interior of the boundary makes it impossible to decompose $W_{m,e_x}$ and $W_{m,e_y}$ into the product of line operators as was the case in $W_{e_x,m}$ and $W_{e_y,m}$. On the other hand, on inspecting Fig. \ref{fig:weg-loop-e}, one realizes that the $Z_0$ operators in the interior cancel out by multiplying $W_{m,e_x}$ and $W_{m,e_y}$. Hence, we can decompose $W_{m,e} \equiv W_{m,e_x}W_{m,e_y}$ as the product of line operators:
\begin{widetext}
\begin{align}
W_{m,e} & = T_m (x_0 \rightarrow x_1 | y_0) T_m (y_0 \rightarrow y_1 | x_1 ) \left[T_m (x_0 \rightarrow x_1 | y_1) \right]^{-1} \left[ T_m (y_0 \rightarrow y_1 | x_0 ) \right]^{-1} . 
\label{eq:weg-loop-m2}
\end{align}
\end{widetext}
Here, the $T_m$ operator is defined by
\begin{align}
& T_m (x_1 \rightarrow x_2 | y) = \prod_{x=x_1\!+\!1}^{x_2}  \left(Z_{2,i} \right)^{x - x_{m0}} \left( Z_{0,i-\hat{x}}\right)^{y_{m0}-y } , \nonumber
\end{align}
\begin{align}
& T_m(y_1 \rightarrow y_2 | x) = \prod_{y=y_1\!+\!1}^{y_2}\left( Z_{1,i} \right)^{y-y_{m0} }  \left(Z_{0,i-\hat{y}} \right)^{x_{m0}-x } . 
\end{align}
The complete cancellation of $Z_0$ takes place for other shapes of the boundary as well, and $W_{m,e}$ can always be decomposed into the product of $T_m$'s. 

\begin{figure*}[tb]
\includegraphics[width=0.8\textwidth]{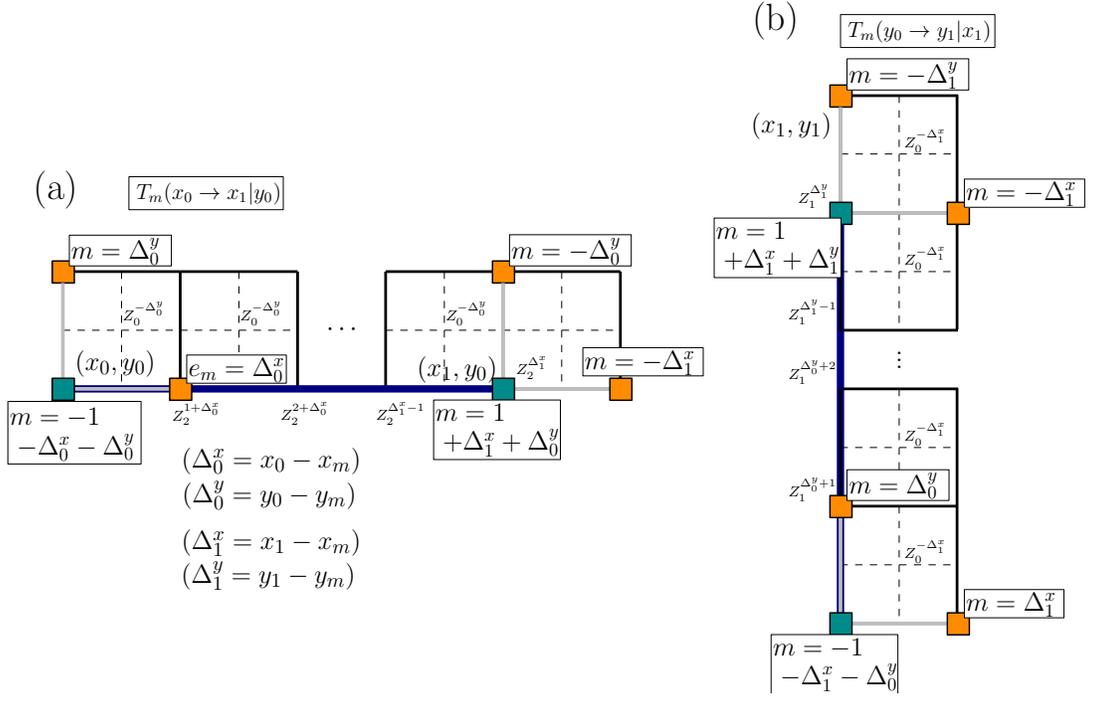}
\caption{Graphical illustration for (a) $T_m(x_0 \rightarrow x_1|y_0)$ and (b) $T_m(y_0 \rightarrow y_1|x_1)$. The navy line represents the domain of $T_x$. The cyan square represents $m$ monopole with charge $\pm 1$ together with other charges composing the auxiliary dipole. The orange square represents the other pair that make up the auxiliary dipole. The auxiliary dipoles are connected by the gray line. 
}
\label{fig:tm}
\end{figure*}

The first operator in $W_{m,e}$ is $T_m (x_0 \rightarrow x_1 | y_0)$, which is illustrated in Fig. \ref{fig:tm}(a). Its role is to translate the $m$ monopoles from $(x_0,y_0)$ to $(x_1,y_0)$ and creating the auxiliary dipoles depending on the modality of the coordinates with respect to $N$. When $y_0-y_{m0} \md N \neq 0$, two auxiliary $m$ dipoles aligned in the $y$-direction are created near $(x_0,y_0)$ and $(x_1,y_0)$, respectively. The dipole near the point $(x_0,y_0)$ has dipole moment $y_0-y_{m0}$, and the other dipole near the point $(x_1,y_0)$ has dipole moment $-(y_0-y_{m0})$. When $x_0-x_{m0} \md N \neq 0$,  $T_m (x_0 \rightarrow x_1 | y_0)$ creates an $x$-directed auxiliary dipole near $(x_0,y_0)$ with the dipole moment $x_0-x_{m0}$. When $x_1-x_{m0} \md N \neq 0$,  $T_m (x_0 \rightarrow x_1 | y_0)$ creates an $x$-directed auxiliary dipole near $(x_1,y_0)$ with the dipole moment is $-(x_1-x_{m0})$. In contrast to $T_x$ or $T_y$ creating the auxiliary dipole only with the $x$- or $y$-component dipole moment, the auxiliary dipole created by $T_m$ has both $x$- and $y$-components and it consists of three monopoles. For example, the auxiliary dipole created by $T_m (x_0 \rightarrow x_1 | y_0)$ near $(x_1,y_0)$ consists of three monopoles with charge $-(y_0-y_{m0})$ at $(x_1,y_0+1)$, charge $-(x_1-x_{m0})$ at $(x_1+1,y_0)$, and charge $(y_0-y_{m0})+(x_1-x_{m0})$ at $(x_1,y_0)$. 

The graphical illustration of $T_m(y_0\rightarrow y_1|x_1)$ is given in Fig. \ref{fig:tm}(b). After the application of $T_m (x_0 \rightarrow x_1 | y_0)$ to the vacuum, applying $T_m(y_0\rightarrow y_1|x_1)$ in succession translates the $m$ monopole with charge $+1$ and the auxiliary dipole from $(x_1,y_0)$ to $(x_1,y_1)$. Subsequently, applying $\left[T_m (x_0 \rightarrow x_1 | y_1) \right]^{-1}$ and $\left[T_m (y_0 \rightarrow y_1 | x_0 ) \right]^{-1}$ moves the $m$ monopoles and the auxiliary dipole from $(x_1,y_1)$ to $(x_0,y_1)$ to $(x_0,y_0)$, which completes the entire loop. In summary, during the operation of $W_{m,e}$, the consecutive $T_m$'s operation braids the $m$ monopole and the accompanied auxiliary dipole around the boundary of the $W_{m,e}$ operator.

\section{Auxiliary dipoles and continuity equations}
\label{sec:aux-dipole}
In this section, we show how to write the charge densities of $e_x$, $e_y$, and $m$ quasiparticles in the presence of their auxiliary dipoles in the continuum theory, and solve the appropriate continuity equations to find the corresponding current densities that match the charge densities. In Sec. \ref{sec:emergent-gauge-derivation} we have shown that in the spin system, the translation of $e_x$, $e_y$, and $m$ monopoles are accompanied by the auxiliary dipoles. The auxiliary dipole moment for $e_x$ ($e_y$) varies in a way that conserves the $y$-component ($x$-component) of the total dipole moment as the $e_x$ ($e_y$) monopole is translated. The auxiliary dipole for $m$ varies in a way that conserves both the $x$- and $y$-component for the $m$ monopole and the accompanying dipole. 

First, consider the $m$ monopole of unit charge moving around by the applications of $T_m$ operators. When the $m$ particle is translated from the initial point ${\bm r}_{m0} = (x_{m0},y_{m0})$ to some point ${\bm r}_m = (x_m,y_m)$, it is accompanied by the auxiliary dipole. The auxiliary dipole, according to the lattice calculation, consists of three $m$ monopoles with charge $-(x_m-x_{m0})$ at $(x_m+1,y_m)$, charge $(x_m-x_{m0})+(y_m-y_{m0})$ at $(x_m,y_m)$, and charge $-(y_m-y_{m0})$ at $(x_m,y_m+1)$. The combined dipole moment of the auxiliary dipole is
\begin{align}
{\bm \mu}^m  = - ( x_m-x_{m0} , y_m -y_{m0} ) ,
\label{eq:aux_dm_em}
\end{align}
which precisely compensates for the dipole moment incurred by the motion of the $m$ particle from ${\bm r}_{m0}$ to ${\bm r}_m$. 

To take the continuum limit we switch the lattice constant from 1 to $\Delta$ and take the limit $\Delta \rightarrow 0$. The positions of three monopoles composing the auxiliary dipole become $(x_m+\Delta,y_m)$, $(x_m,y_m)$, and $(x_m,y_m+\Delta)$. The monopole charges are modified accordingly to $-\frac{1}{\Delta}(x_m-x_{m0})$, $\frac{1}{\Delta}(x_m-x_{m0})+\frac{1}{\Delta}(y_m-y_{m0})$, and $-\frac{1}{\Delta}(y_m-y_{m0})$, respectively.  We can write the charge density for the auxiliary dipole as
\begin{widetext}
\begin{align}
\rho^{{\bm \mu}^m}({\bm r}) =& \frac{\left(x_m\!-\!x_{m0}\right)}{\Delta} \Big[ -\! \delta\left(x\!-\!x_m\!-\!\Delta\right)\delta\left( y\!-\!y_m\right) \!+\! \delta\left(x\!-\!x_m \right)\delta\left( y\!-\!y_m\right) \Big] \nn
&+\!
\frac{\left(y_m\!-\!y_{m0}\right)}{\Delta}\Big[ -\!\delta\left(x\!-\!x_m\right)\delta\left( y\!-\!y_m\!-\!\Delta\right) 
\!+\!\delta\left(x\!-\!x_m\right)\delta\left( y\!-\!y_m\right)  \Big], 
\label{eq:rho-m-1}
\end{align}
\end{widetext}
which becomes, in the $\Delta\rightarrow0$ limit, 
\begin{align}
\rho^{{\bm \mu}^m}({\bm r}) =& (x_m -x_{m0}) \partial_x \delta(x-x_m ) \delta(y-y_m )  \nn 
& + (y_m -y_{m0}) \delta(x-x_m ) \partial_y \delta(y-y_m ) \nn 
=&   ( {\bm r}_m - {\bm r}_{m0} ) \cdot {\bm \nabla}  \delta^2({\bm r}-{\bm r}_m ),
\end{align}
where $\delta^2({\bm r}-{\bm r}_m ) =  \delta(x-x_m ) \delta(y-y_m )$.
Together with the original $m$ monopole density $\rho^{m,0} ({\bm r}) =\delta^2 ({\bm r} - {\bm r}_m )$, the net density becomes
\begin{align}
\rho^{m}({\bm r}) =&  \rho^{m,0}({\bm r})  + \rho^{{\bm \mu}^m}({\bm r}) \nn 
 =&  \delta^2({\bm r}-{\bm r}_m )+ ( {\bm r}_m - {\bm r}_{m0} ) \cdot {\bm \nabla}   \delta^2({\bm r}-{\bm r}_m ). 
\end{align}

Similar consideration applies to the two electric monopoles. When the $e_x$ monopole is translated from ${\bm r}_{x0} = (x_{x0},y_{x0})$ to ${\bm r}_x = (x_x, y_x)$, it is accompanied by an auxiliary dipole consisting of $e_x$ monopole with charge $(x_x-x_{x0})$ at $(x_x,y_x-1)$ and charge $-(x_x-x_{x0})$ at $(x_x,y_x)$. By following the same procedure for taking the continuum limit as before, we obtain the charge density of $e_x$ particle as
\begin{align}
\rho^x = & \rho^{x,0} + \rho^{\mu^x} = \delta^2({\bm r}-{\bm r}_x) + (y_x-y_{x0}) \partial_y \delta^2({\bm r}-{\bm r}_x) ,  \nn 
\rho^y = &  0,
\end{align}
where $\rho^{x,0} ({\bm r}) = \delta^2 ({\bm r}-{\bm r}_x)$ is the $e_x$ monopole density, and $\rho^{\mu^x}$ is the density of the auxiliary dipole. Note that we need to consider the density of the $e_y$ monopole simultaneously, albeit $\rho^y=0$, since it is coupled to $\rho^x$ by the continuity equations in Eq. (\ref{eq:continuity}). For the motion of $e_y$ we find 
\begin{align}
\rho^y = & \rho^{y,0} + \rho^{\mu^y} = \delta^2({\bm r}-{\bm r}_x) + (x_y -x_{y0}) \partial_x \delta^2({\bm r}-{\bm r}_x) , \nn 
\rho^x = & 0. 
\end{align}

\bibliography{sc}
\end{document}